\documentclass[iop]{emulateapj}

\usepackage{epstopdf}
\usepackage[utf8]{inputenc}
\usepackage{hyperref}
\usepackage{mathtools}
\usepackage{amsmath}

\hypersetup{colorlinks=True,linkcolor=magenta,citecolor=blue,filecolor=cyan,urlcolor=red}

\newcommand{\src}{G12v2.43\,}
\newcommand{\mgf}{$\rm(\frac{10}{\mu})$}
\newcommand{\lsol}{$\,\rm{L_{\odot}}\,$}
\newcommand{\msol}{$\,\rm{M_{\odot}}\,$}

\shorttitle{Young Stellar Population in a z $\sim$ 3 HyLIRG}
\shortauthors{Vishwas et al.}
\slugcomment{draft version February 26, 2018, accepted for publication in The Astrophysical Journal}

\begin{document}

\title{Detection of [O\,{\sc iii}] at \lowercase{z}$\sim$3: A Galaxy above the Main-Sequence, rapidly assembling its Stellar Mass}

\author{Amit Vishwas\altaffilmark{1}, Carl Ferkinhoff\altaffilmark{2}, Thomas Nikola\altaffilmark{3},  Stephen C. Parshley\altaffilmark{3},  Justin P. Schoenwald\altaffilmark{3}, \\Gordon J. Stacey\altaffilmark{1}, Sarah J.U. Higdon\altaffilmark{4},  James L. Higdon\altaffilmark{4}, Axel Wei{\upshape{\ss}}\altaffilmark{5}, Rolf G{\"u}sten, \altaffilmark{5}, and Karl M. Menten\altaffilmark{5}}

\email{vishwas@cornell.edu}
\altaffiltext{1}{Department of Astronomy, Cornell University, Ithaca, NY 14853, USA.}
\altaffiltext{2}{Department of Physics, Winona State University, Winona, MN, 55987, USA.}
\altaffiltext{3}{Cornell Center for Astrophysics and Planetary Science, Cornell University, Ithaca, NY 14853, USA.}
\altaffiltext{4}{Department of Physics and Astronomy, Georgia Southern University, Statesboro, GA 30460, USA.}
\altaffiltext{5}{Max-Planck-Institut f{\"u}r Radioastronomie, Auf dem H{\"u}gel 69, 53121 Bonn, Germany.}

\begin{abstract}
We detect bright emission in the far infrared fine structure [O\,{\sc iii}] 88\,$\mu$m line from a strong lensing candidate galaxy, H-ATLAS\,J113526.3-014605, hereafter \src, at z = 3.127, using the $\rm 2^{nd}$ generation Redshift (z) and Early Universe Spectrometer (ZEUS-2) at the Atacama Pathfinder Experiment Telescope (APEX).
This is only the fifth detection of this far-IR line from a sub-millimeter galaxy at the epoch of galaxy assembly. The observed [O\,{\sc iii}] luminosity of $7.1\times10^{9}\,$\mgf\lsol likely arises from H{\sc ii} regions around massive stars, and the amount of Lyman continuum photons required to support the ionization indicate the presence of $(1.2-5.2)\times10^{6}\,$\mgf\, equivalent O5.5 or higher stars; where $\mu$ would be the lensing magnification factor. The observed line luminosity also requires a minimum mass of $\sim 2\times 10^{8}\,$\mgf\msol in ionized gas, that is $0.33\%$ of the estimated total molecular gas mass of $6\times10^{10}\,$\mgf\msol. We compile multi-band photometry tracing rest-frame UV to millimeter continuum emission to further constrain the properties of this dusty high redshift star-forming galaxy. Via SED modeling we find \src is forming stars at a rate of 916\,\mgf\msol$\rm{yr^{-1}}$ and already has a stellar mass of $8\times 10^{10} $\mgf\msol. We also constrain the age of the current starburst to be $\leqslant$\,5 million years, making \src a gas rich galaxy lying above the star-forming main sequence at z$\sim$3, undergoing a growth spurt and, could be on the main sequence within the derived gas depletion timescale of $\sim$66 million years.
\end{abstract}

\keywords{galaxies: high-redshift galaxies: starburst  galaxies: stellar content galaxies: individual (G12v2.43) instrumentation: spectrographs submillimeter: galaxies }

\thanks{\emph{APEX} is a collaboration between the Max-Planck-Institut f{\"u}r Radioastronomie, the European Southern Observatory, and the Onsala Space Observatory.}


\section{Introduction} \label{sec:intro}

Over the past 20 years, wide-field multi-band surveys have demonstrated that the star formation rate per unit comoving volume of the Universe rose quickly soon after re-ionization and peaked at redshifts between z $\sim$ 3 and 1 (look-back times of $\sim$11.5 to 7.7 Gyr) at rates 10 to 15 times the present-day values \citep[see e.g.,][]{Madau2014}. Locally, and even back beyond redshift 3, a substantial fraction of star formation within galaxies is obscured by dust. This dust absorbs starlight, and re-radiates its power in the far-infrared continuum. For most high luminosity star-forming galaxies, the far-infrared luminosity exceeds the optical/UV luminosity so we have come to call these dusty star-forming galaxies (DSFGs). DSFGs dominate the rise in star formation rate density looking back in time to at least beyond redshift of 3, so it is important to study DSFGs in their rest-frame far-infrared bands to properly understand the history of star formation in the Universe.

We have constructed two sub-millimeter (submm) grating spectrometers, ZEUS \citep{Hailey-Dunsheath2009} and ZEUS-2 \citep{Ferkinhoff2014}, in order to measure far-infrared fine-structure line emission from luminous star-forming galaxies between z=1-5. These far-infrared lines (e.g. [C\,{\sc ii}] 158, [N\,{\sc ii}] 122 \& 205, [O\,{\sc iii}] 88 \& 52\,$\mu$m) are important coolants of the gas, and excellent probes of both the physical properties of the emitting medium and the dominant sources of luminosity -- a burst of star formation activity or accretion onto a super-massive black hole. They have advantages over the optical lines in that they are typically optically thin, insensitive to extinction by dust\footnote{Extinction optical depth of unity requires a gas column of $\rm N_{H}>10^{24}\,\rm{cm^{-2}}$ for $\lambda\geq$\,60$\mu$m \citep{Draine2003}.}, and for lines arising from ionized gas, they are also insensitive to the ionized gas temperature. In the local Universe, a full complement of far-IR fine structure lines have been studied in many sources, helping us to constrain the properties of the interstellar medium and the host stellar populations, using airborne and space based instruments \citep[e.g.,][]{Brauher2008, Gracia-Carpio2011, Farrah2013, Cormier2015, Diaz-Santos2017}. At z$>$0.2, sensitivity of instruments and the Earth’s atmosphere makes measurements of these lines very challenging. Using our ZEUS instruments, we are surveying star-forming galaxies in the redshift 1 to 5 epoch in their far-infrared fine-structure line emission, including the [C\,{\sc ii}] 158\,$\mu$m line \citep[]{Hailey-Dunsheath2010, Stacey2010, Ferkinhoff2014, Brisbin2015}.

The far-IR lines of [O\,{\sc iii}] and [N\,{\sc ii}] arise in H{\sc ii} regions, and are prominent coolants tracing  the physical conditions and excitation mechanism of gas in sites of active star formation. The lines individually allow us to measure the flux of ionizing radiation while, the [O\,{\sc iii}] 88\,$\mu$m/[N\,{\sc ii}] 122\,$\mu$m line ratio (modulo abundance considerations) is primarily sensitive to the hardness of the radiation field \citep[e.g.][]{Rubin1985}. Combined, these constraints provide a luminosity-weighted measurement of the number and type of the of the most massive stars still on the main sequence and hence the intensity and age of the most recent starburst \citep[c.f.][]{Ferkinhoff2010}. To pursue our goal of characterizing the starbursts in high redshift DSFGs, we are surveying the [O\,{\sc iii}] 88\,$\mu$m and [N\,{\sc ii}] 122$\mu$m lines using our submm grating spectrometers ZEUS (on CSO) and ZEUS-2 (on APEX). Theoretical modeling of emission regions and observations of local galaxies suggest that the [N\,{\sc ii}] 122$\mu$m line is fainter than the [O{\sc iii}] 88\,$\mu$m line, and only two detections have been reported at z$>$1, both in composite starburst-AGN systems, by \citep[]{Ferkinhoff2011,Ferkinhoff2015}. To date, the [O\,{\sc iii}] 88 $\mu$m line has only been reported in five high redshift submm galaxies (SMGs); two detected with ZEUS, two with Herschel-SPIRE in z$\sim$3 lensed SMGs \citep[]{Valtchanov2011, Rigopoulou2018} and one with ALMA at z$\sim$6.9 \citep[]{Marrone2017}. Three additional detections of the [O\,{\sc iii}]\,88$\mu$m line have been reported in clumps associated with Ly-$\alpha$ systems at z$\sim$7-8, using ALMA \citep[]{Inoue2016, Carniani2017, Laporte2017}. Here, we report observations of the [O{\sc iii}] 88\,$\mu$m line in a Herschel discovered SMG, \src at z=3.127 with ZEUS-2 on the APEX telescope in the 350\,$\mu$m waveband.

\src (RA: 11:35:26.3 Dec: -01:46:06.5, J2000) was discovered in the Herschel Astrophysical Terahertz Large Area Survey \citep[H-ATLAS,][]{Clements2010}, and was selected as a candidate high redshift gravitationally lensed source due to its large 500 $\mu$m flux density (S$_{500\mu m}>$0.1 Jy). It was confirmed as a high redshift (z=3.1276$\pm$0.0005) system through mm–band spectroscopy using the Green Bank Telescope \citep[GBT,][]{Harris2012} and the Northern Extended Millimeter Array \citep[NOEMA,][]{Yang2016}. These observations show powerful emission in low–J CO and H$_{\rm{2}}$O rotational lines, thereby confirming \src’s extreme luminosity. In a high spatial resolution study, using the Sub–millimeter Array (SMA), \citet{Bussmann2013} reported the detection of (observed frame) 896 $\mu$m continuum. The source was marginally resolved but did not show any extended emission or obvious signatures of gravitational lensing like multiple images or lensing arcs at 0\farcs8 (FWHM) scale. Follow up deep near-infrared imaging to find the foreground lensing galaxy by \citet{Calanog2014} detected no significant emission in the K$_s$ band either from the foreground lens or the background high redshift galaxy. We note that the co-ordinates reported above are the true centroid of the emission seen in the SMA map and are offset by 1\farcs4 from the those reported in \citet{Bussmann2013}.

A reasonable explanation for its extreme luminosity, L$_{\rm{IR}}$, of $\sim1.2\times10^{14}\,$\lsol \citep{Bussmann2013} could be magnification due to gravitational lensing, as was suggested by \citet{Harris2012} due to the large CO luminosity and small line FWHM. For discussion related to physical quantities in this paper, we adopt a scaling factor of \mgf, where $\mu$ would be the true lensing magnification factor. Within the purview of existing observations, the influence of gravitational lensing for \src\,remains unconstrained. However as a reference for the reader, galaxies with such extreme observed luminosities and well constrained lensing models have a median magnification factor of $\mu\sim$\,6 \citep[e.g.,][]{Bussmann2013, Spilker2016}. If confirmed, the apparent brightness of \src would allow future observations to study a normal galaxy building up its stellar mass at sub-kpc resolution and study the interplay between star formation and galaxy evolution only 2 billion years after the Big Bang. 


In this paper we present the [O\,{\sc iii}] 88\,$\mu$m line observations and compile broadband photometry from UV to mm wavelengths towards \src to study the properties of its ionized gas, dust and stellar populations. In Section \ref{sec:Z2}, we present the ZEUS-2/APEX observations and discuss the constraints on its young stellar population and ionized gas mass. In Section \ref{sec:other_obs}, we compile new observations of \src from various archives; namely near-infrared data taken with the Hubble and Spitzer Space Telescopes and far-infrared photometry and spectroscopy with the Herschel Space Observatory. In Section \ref{sec:Dis}, we present spectral energy distribution modeling of the photometric data and discuss the ISM properties of \src and the potential of it being a gravitationally lensed system. Throughout this paper we assume a flat $\Lambda$CDM cosmology with a Hubble constant of 70 $\rm{km\,s^{-1} Mpc^{-1}}$, $\Omega_{M}$= 0.3, and $\Omega_{\Lambda}$=0.7, giving \src a luminosity distance of 26.7 Gpc and a linear scale of 7.6 kpc arcsec$^{-1}$.

\section{ZEUS-2/APEX Observations and results} \label{sec:Z2}

The second generation redshift(z) and Early Universe Spectrometer (ZEUS–2) is a grating spectrometer optimized for detecting broad (few 100 km\,s$^{-1}$) spectral lines from distant galaxies as they are redshifted into the short sub–mm telluric windows \citep[][]{Ferkinhoff2012, Parshley2012, Ferkinhoff2014}. We observed \src with ZEUS-2 at the Atacama Pathfinder Experiment (APEX) telescope \citep{Guesten2006} in 2014 October under very good weather conditions. The telescope was pointed at the co-ordinates reported in \citet{Bussmann2013}, offset by 1\farcs4 from the peak of the sub-mm emission, but well within the ZEUS-2/APEX beam. The precipitable water vapor remained stable between 0.5 and 0.56 mm which corresponds to a line of sight transmission of 31-42\% at 365\,$\mu$m during the course of the observations.

\begin{figure}[!tbh]
\epsscale{0.8}
\plotone{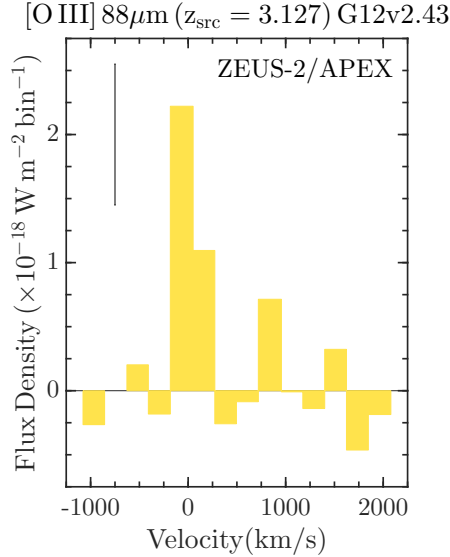}
\caption{ZEUS-2/APEX spectrum of the [O{\sc iii}]\,88\,$\mu$m emission line observed in \src at a redshift of z = 3.127. The velocity scale is with respect to the source redshift. Typical per bin 1-$\sigma$ error is shown in top-left.
\label{fig:Z2OIII}}
\end{figure}

The [O\,{\sc iii}] 88\,$\mu$m line was observed at 364.7 $\mu$m based on the redshift reported by \citet{Harris2012}. The resolving power of the instrument at 365\,$\mu$m (in the 5th order of the echelle grating) is R$\sim$960. Each spectral pixel covers $\sim$313\,km\,s$^{-1}$ in velocity space with the array providing an instantaneous coverage of $\sim$3500\,km\,s$^{-1}$. Data was taken in standard chop/nod mode, with 2 Hz chop frequency and a 30\arcsec \,azimuthal chop-throw. The source elevation was between 50-68$\degr$ during the observations. Three grating settings were used to move the line along spectral pixels to increase total spectral coverage and eliminate gaps due to non–functioning pixels. As a result, the data are sampled at 230\,km\,s$^{-1}$, finer than our resolution element of 313\,km\,s$^{-1}$. We obtained 8$\times$10 minute integrations with a chop efficiency of 63\% giving a total on-source integration time of 25.2 minutes. Pointing and focus observations were taken on the $^{12}$CO(6-5) line from IRC10216, and were repeated every 20 minutes. Pointing was found accurate to within 2\arcsec \,over the duration of the observations. The ZEUS-2/APEX beam at 365\,$\mu$m was measured using Uranus as 7\farcs8$\pm$0$\farcs$9. Gain calibration for spectral pixels was done by chopping an ambient temperature blackbody against cold sky before and after each set of on–source observations and flux calibration was verified with Jupiter and Uranus. The signal we measured from Uranus, indicates that the point source coupling for APEX is $\sim$40\% at 365\,$\mu$m. We detected the [O\,{\sc iii}] 88\,$\mu$m line, shown here in Figure \ref{fig:Z2OIII}, at a flux level of $3.2\pm0.4\times10^{-18}\,$Wm$^{-2}$ (S/N=7.5), or equivalently 116$\pm$15 Jy km\,s$^{-1}$, with an estimated calibration uncertainty of 30\%. We advise the reader to use the velocity FWHM of 225\,km\,s$^{-1}$, derived by spectrally resolved observations of \citet{Harris2012} and \citet{Yang2016} for the CO and H$_2$O lines respectively, to estimate the peak line flux density.

\subsection{{\rm [O\,{\sc iii}]} line emission: Young stars and Ionized gas mass} \label{subsec:OIII_prop}

As the O$^{{\rm++}}$ ion requires 35\,eV to form, it indicates the presence of a very hard ionizing source which could either be upper main sequence stars, with effective temperature $>$36,000\,, in the vicinity of the emitting region or a nearby Active Galactic Nucleus (AGN). Assuming the star formation dominated scenario, the [O\,{\sc iii}] line emission can be used to constrain the type of stars producing the ionizing radiation (O8 or hotter) and the line luminosity can be used to estimate the flux of ionizing photons required to support the observed emission. Using theoretical models for upper main sequence stars, we can scale the number of ionizing photons produced by a certain spectral type of O-star to estimate the number of such stars present, on average, in the host galaxy. Also, since the most massive stars spend a relatively short  amount of time on the main sequence, up to about 3–10 million years,  the observed [O\,{\sc iii}] line emission places a constraint on the age of the most recent starburst.

Our observations of the [O\,{\sc iii}] 88\,$\mu$m line flux corresponds to a luminosity of $7.1\times10^{9}$\mgf\lsol at the distance of \src. In Section \ref{subsec:sed}, we suggest that the emission seen from G12v2.43 is dominated by star-formation activity and not from an AGN with the help of broadband SED modeling. With the [O\,{\sc iii}] line flux, we can constrain the number of photons capable of doubly ionizing oxygen, and estimate the number of O-stars given an upper mass cutoff. We use the H{\sc ii} region models of \citet{Rubin1985} to scale the observed line flux and estimate the number of Lyman continuum photons. The effective stellar temperatures, T$_{\rm eff}$, used in the models are matched with those of the spectral type of main sequence O–stars using the calibration of \citet{Martins2005}. The main sequence lifetime is based on the Hydrogen burning timescale reported by \citet{Ekstrom2012} for massive stars of solar metallicity. We use Rubin’s ‘K’ models which are based on stellar atmosphere models from \citet[][]{Kurucz1979, Kurucz1993} with an [O/H] abundance of $6.76\times10^{-4}$, to predict the [O\,{\sc iii}] line luminosity as a function of the effective stellar temperature of the star (T$_{\rm eff}$=31-45 kK), the Lyman continuum (LyC) photon rate (Q$_{0}$=10$^{49-50}$\,s$^{-1}$), and the electron number density in the H{\sc ii} regions 
(n$_{\rm e}=10^{2-4}\,$cm$^{-3}$). The models are set up such that the line intensities scale linearly for models with varying Q$_{0}$, so that the derived number of stars for the model with Q$_{0}=10^{49}\,$s$^{-1}$ would be 10 times the number of stars derived for models with Q$_{0}=10^{50}\,$s$^{-1}$. The models with n$_{\rm e}=10^{3}\,$cm$^{-3}$ and T$_{\rm eff}$=40,000 and 45,000K provide the best fit with the derived number of LyC photon rate, Q$_{0}\approx 5.7\times10^{55}$\,\mgf\,s$^{-1}$. This LyC flux level require $(1.2-5.2)\times10^{6}$\mgf\, O3V to O5.5V stars. An estimate using the total bolometric luminosity of such upper main sequence stars present in the galaxy suggests that if all their starlight were absorbed by dust and re-radiated in the far-infrared, that could account for $(0.8-1.3)\times10^{12}$\mgf\lsol, i.e., $\sim$12\% of the observed FIR luminosity of $8.3\times10^{12}$\mgf\lsol.

Following \citet{Ferkinhoff2010}, in the high density, high temperature limit, we can estimate the minimum mass ionized nebula required to support the luminosity of the observed [O\,{\sc iii}] line as:

\begin{equation}
M_{\rm min}^{\rm H^+} = 4\pi d_{\rm L}^{2}\,\frac{{\rm F_{O\,III}}}{\frac{g_l}{g_t} A_{ul} h \nu_{ul}}\,\frac{m_{\rm H}}{\xi_{\rm O^{++}}}
\end{equation}

Here, F$_{\rm O\,III}$ is the observed line flux (W\,m$^{-2}$), d$_{\rm L}$ is the luminosity distance (m), g$_{l}$ is the statistical weight (2J+1) for the J=1 emitting level,  g$_{t}$=$\sum\nolimits_{i} g_{i}\,e^{-\Delta E_{i}/kT}$, is the partition function, A$_{ul}$ is the spontaneous emission coefficient (s$^{-1}$), h is the Planck’s constant (J-s), $\nu_{ul}$ is the rest frequency of the line, 3393.00624\,GHz (88.356\,$\mu$m), m$_{\rm H}$ is the mass of a hydrogen atom, and $\xi_{\rm O^{++}}$, is the relative abundance of (${\rm O^{++}}/{\rm H^{+}}$). Adopting a nebular gas phase abundance, [O/H]=$5.9\times10^{-4}$ and assuming all the oxygen is doubly ionized ([O/H]=[${\rm O^{++}}/{\rm H^{+}}$]), we find that the minimum ionized gas mass in \src is $(2.0\pm 0.3)\times10^{8}$\mgf\msol. The minimum mass of doubly ionized oxygen itself is, $\sim1.9\times10^{6}$\mgf\msol.

\section{Supporting Observations} \label{sec:other_obs}

Previous observations of \src have been reported by \citet{Harris2012}, \citet{Bussmann2013}, and \citet{Yang2016}. However, these authors only discuss emission at wavelengths longer than (observed frame) 200\,$\mu$m, so that many of the source properties, including the stellar mass and the total luminosity are not well constrained. We compile published and archival observations to produce a rest-frame near–UV to mm SED which we use to constrain the star formation history, stellar mass, total luminosity and dust properties. As the source is compact and the beam of the various instruments vary from 0\farcs15-35\arcsec\,, flux densities (or limits) were derived via aperture photometry in the recommended manner for each instrument assuming a point source. These data are summarized in Table \ref{tab:phot}.

\subsection{Archival: Hubble Space Telescope WFC3} \label{subsec:HST}

As part of a snapshot program to identify gravitationally lensed galaxies (PI: Negrello, ID: 12488), \src was observed with the HST Wide-field Camera using the wide near-infrared filter F110W for 711.7 seconds in 2013 July. In a lensing system, the near–infrared images are typically used to identify the foreground lensing galaxy and the rest frame near–UV to optical light tracing stellar emission in the background higher redshift galaxy. We obtained the pipeline calibrated images from the Hubble Legacy Archive to investigate the presence of either a foreground galaxy, or any structure like arcs or an Einstein ring, that could be characteristic of strong gravitational lensing.

\begin{figure}[!t]
\plotone{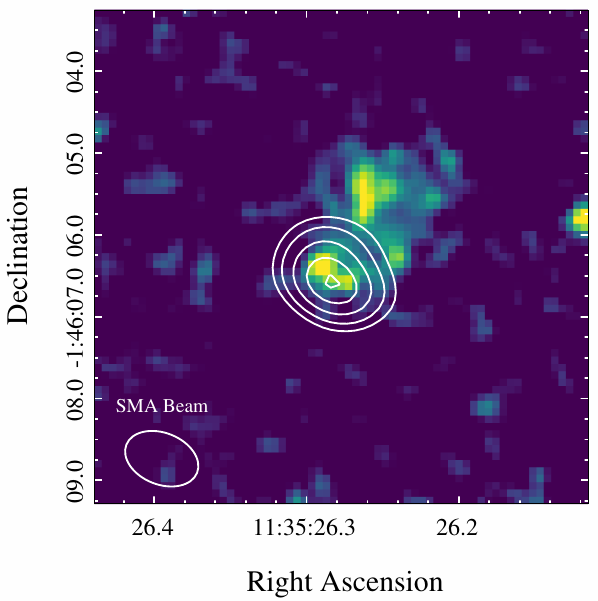}
\caption{(Background: Color map)- A 6\arcsec$\times$6\arcsec\, stamp of 1.1\,$\mu$m WFC3 image at the position of \src. The emission in the HST/WFC3 image shows two sources near the expected high-z source location, separated by $\sim$1\arcsec . (Foreground, White Contours) 896\,$\mu$m continuum emission detected at the position of \src with the SMA \citep[from][]{Bussmann2013}. The contours, starting from the center, show the peak of the emission detected at 24-$\sigma$ and moving out to (17, 12, 8.5, 6)-$\sigma$ levels. The white ellipse in the bottom-left corner indicates the beam size of the SMA observations (FWHM: 0\farcs93$\times$0\farcs63).
\label{fig:hstsma}}
\end{figure}

We identify a source in the near-IR WFC3 F110W image at 5.1–$\sigma$ significance that is consistent with the centroid of the emission seen from the high-z galaxy in interferometric observations from SMA (Figure \ref{fig:hstsma}). The emission is faint (m$_{\rm AB}$=24.2), and another source is seen 1\arcsec\,north-west of the location of the high-z galaxy. Previous attempts at identifying the foreground lensing galaxy in the K$_{s}$ band using deep Keck observations have been unsuccessful \citep{Calanog2014} and, we cannot say with certainty whether these two sources are actually patchy emission in rest-frame near-UV from \src or could be partially attributed to a foreground, perhaps lensing galaxy. As the emission is extended around the two features, to estimate the flux density of the high-z source, we perform photometry with a 1\arcsec\, circular aperture centered at the peak of the SMA emission to avoid picking up flux from the nearby source and subtract the median sky from each pixel.

\subsection{Archival: Spitzer Space Telescope}

We obtained the pipeline calibrated images of \src from the Spitzer Heritage Archive. The source was observed using the Spitzer Infrared Array Camera (IRAC) in Cycle 11 (PI: A. Cooray, Program ID 80156) in the 3.6 and 4.5\,$\mu$m bands. The on-source integration time was 706 seconds. Emission in the IRAC bands traces near-IR light from the high redshift source at the effective rest-frame wavelength of 872\,nm and 1090\,nm. We find bright emission in both bands at the source location, consistent with the HST F110W image reported in the previous section and with the interferometric SMA imaging, and no signs of extended emission or artifacts. We perform photometry on the Level–2 pipeline products (post–BCD) following the methodology described in the Appendix B of the IRAC instrumentation handbook. Since the data are undersampled and there is no clear evidence for extended emission at the IRAC resolution, we perform source fitting and aperture photometry in the recommended manner using both a 3.6\arcsec\,and 6\arcsec\,aperture to estimate the flux density. The uncertainties are estimated from the uncertainty image provided with the science data products from the archive and compared to the background estimated using a large annulus around the source. The uncertainty derived from the uncertainty maps is smaller than the 5\% calibration accuracy of IRAC, but the difference between various size apertures is larger, about 10\%. We quote this 10\% uncertainty for both of the measured flux densities for \src.

\subsection{Archival: Wide–field Infrared Survey Explorer}

We utilized the publicly available All-sky data release from the Wide-field Infrared Survey Explorer (WISE) \citep{Wright2010} to look for emission from \src between 3.4 and 22\,$\mu$m. We queried the AllWISE point source catalog for entries within 5\arcsec\, of the Spitzer position. The source was only detected (5$\sigma$) in W1 (3.35\,$\mu$m) band and we derived 3 sigma upper limits for the flux density in the W2 (4.6\,$\mu$m), W3 (11.56\,$\mu$m) and W4 (22.1\,$\mu$m) bands based on the magnitudes reported in the catalog, see Table \ref{tab:phot}.

\setlength{\tabcolsep}{3pt}
\begin{deluxetable}{cccc}[!b]
	\tabletypesize{\scriptsize}
	\centering
	\tablewidth{0pt}
\tablecaption{Photometry Data \label{tab:phot}}
\tablehead{
\colhead{Wavelength} & \colhead{Frequency} & \colhead{Flux Density} & \colhead{Instrument} \\
\colhead{($\mu$m)} & \colhead{(GHz)} & \colhead{(mJy)} & \colhead{}
}
\startdata
1.15 & 260689.1 & 0.9 $\pm$ 0.2$\times10^{-3}$ & HST/WFC3 \\
3.4 & 88174.3 & 29 $\pm$ 6$\times10^{-3}$ & WISE/W1 \\
3.6 & 83275.7 & 31 $\pm$ 3$\times10^{-3}$ & Spitzer/IRAC1 \\
4.5 & 66620.5 & 45 $\pm$ 4$\times10^{-3}$ & Spitzer/IRAC2 \\
4.6 & 65172.3 & 31 $\pm$ 11$\times10^{-3}$ & WISE/W2 \\
12 & 24982.7 & $<$ 0.49 & WISE/W3 \\
22 & 13626.9 & $<$ 3.9& WISE/W4 \\
70 & 4282.7 & 19 $\pm$ 3 & Herschel/PACS \\
100 & 2997.9 & 56 $\pm$ 6 & Herschel/PACS \\
160 & 1873.7 & 180 $\pm$ 12 & Herschel/PACS \\
250 & 1199.2 & 296 $\pm$ 17 & Herschel/SPIRE \\
350 & 856.5 & 306 $\pm$ 24 & Herschel/SPIRE \\
500 & 599.6 & 214 $\pm$ 23 & Herschel/SPIRE \\
896 & 334.6 & 50 $\pm$ 3 & SMA\tablenotemark{a} \\
1064 & 281.8 & 36.4 $\pm$ 0.3 & NOEMA\tablenotemark{b} \\
1252 & 239.4 & 22.5 $\pm$ 0.5 & NOEMA\tablenotemark{b} \\
208900 & 1.435 & $<$ 0.36 & VLA/FIRST \\ 
\enddata
\tablenotetext{a}{The SMA photometry was previously presented in \citet{Bussmann2013}, note change in center wavelength from 880\,$\mu$m to 896\,$\mu$m}
\tablenotetext{b}{The NOEMA mm-wave photometry is  adopted from \citet{Yang2016}}
\tablecomments{Wavelength/Frequency in observed frame; All upper limits are 3-$\sigma$}
\end{deluxetable}

\pagebreak
\subsection{Archival: Herschel Space Telescope}

\subsubsection{Photometry}

Due to the redshift of \src (z=3.127), the mid-infrared part of the spectral energy distribution, which could help constrain emission from a hot dust component or an obscured AGN, is shifted into the far-infrared bands covered by the PACS and SPIRE instruments on-board the Herschel Space Observatory.

We present observations of \src taken in the 70, 100, and 160\,$\mu$m bands of the PACS photometer \citep{Poglitsch2010}. The observations used here were taken as a part of the observing programs, OT1\_rivison\_1, Observation ID: 1342224173,74, on OD 792 and OT2\_jwardlow\_2, Observation ID: 1342257109-112, on OD 1309. The source was observed for 276–558 seconds in each band. This data was processed using HIPE version 15 \citep{Ott2010} through pipeline version 14.2. We combined all the observations of \src in each PACS band and then perform point-source aperture photometry using the task, sourceExtractorSussextractor. We detect the source in all three PACS bands. The flux density derived in the 160\,$\mu$m band is lower by 17\% than the value reported by \citet{Wardlow2017} but consistent with that reported in the PACS point source catalog \citep{Marton2017}. Statistical error in the measurement was calculated by estimating the noise level in the map in each band. The SPIRE bands span across the peak of the dust emission at z$\sim$3 and are critical to estimate dust properties and the far-infrared luminosity. We use the flux density reported in the SPIRE point source catalog \citep{Schulz2017} at the location of \src. The color correction required for SPIRE data points is 1.02 at 250\,$\mu$m, 0.97 at 350\,$\mu$m and 0.95 at 500\,$\mu$m. In Table \ref{tab:phot}, the errors reported with the SPIRE flux densities are the confusion noise in the maps as reported by the point source catalog. The confusion noise is much larger than the 5\% calibration uncertainty or the statistical noise derived by the sourceExtractorTimeline task in HIPE (2-3\%).

\subsubsection{Spectroscopy}\label{subsec:SPIRE_spec}

\src was observed with the SPIRE Fourier Transform Spectrometer \citep{Griffin2010}. The observations used here were taken as a part of the observing program OT1\_rivison\_1, Observation ID: 1342247744, on OD 1150 in high-resolution mode for 13752 seconds towards the end of the Herschel mission. We reprocessed the data through HIPE 15 with SPIRE calibration version 14.3 and corrected for instrumental artifacts. The spectral shape and absolute flux calibration were verified by comparing the SPIRE photometry with synthetic measurements derived from the corrected spectral data, using the task spireSynthPhotometry within HIPE 15. The inherent instrument response of a Fourier Transform Spectrometer is a sinc function, that has 20\% sidelobes associated with each peak. Instead of looking at a single co–added scan and deriving limits for the line fluxes, we improve the noise characterization by creating multiple realizations of averaged scans by randomly selecting 100 out of the 200 available scans. As only half of the available scans were used in each realization, the sensitivity in each scan would, in principle be worse by a factor of $\sqrt{2}$. From these realizations, we estimate the noise in a 5\,GHz band centered at the frequency of individual far-IR lines. No lines are detected at high significance and the 3-$\sigma$ limits are listed in Table \ref{tab:fsl}.

The electronic ground state of doubly ionized oxygen is split by fine structure interactions into three levels, ground ($^{3}P_{0}$), and two excited states, $^{3}P_{1}$, and $^{3}P_{2}$. As the fine-structure states are only a few hundred K above the ground state, ions can be collisionally excited by free electrons in the H{\sc ii} regions (typical temperature $\sim$ 8000\,K), to occupy these states. The 3-$\sigma$ limit for the upper transition, $^{3}P_{2}\rightarrow\,^{3}P_{1}$ at 52\,$\mu$m, of L$_{\rm [O\,III]}<\,6.1\times10^{-18}$\,W\,m$^{-2}$, in conjunction with our detection of the ground state transition, $^{3}P_{1}\rightarrow\,^{3}P_{0}$ at 88\,$\mu$m allows us to constrain the density of the emitting gas by comparing the observed line ratio to the theoretical line emissivity ratio \citep[][]{Rubin1989}. The luminosity ratio of the two [O{\sc iii}] lines, \(\frac{\rm L_{[O\,III]\,52\,\mu m}}{\rm L_{[O\,III]\,88\,\mu m}}<1.92\), suggests that the emitting gas has density, n$<610\,$cm$^{-3}$. This is consistent with our choice of H{\sc ii} region models with density, n=$10^{2-3}\,$cm$^{-3}$, used to interpret the [O{\sc iii}]  88\,$\mu$m line emission in Section \ref{sec:Z2}. Similarly, the 3-$\sigma$ limit for the the [C\,{\sc ii}] 158\,$\mu$m line flux, ${\rm L_{C\,II}}<1.4\times10^{-17}\,$W\,m$^{-2}$, yields a [C\,{\sc ii}]/FIR luminosity ratio $\leq $0.4\%, consistent with the ratio observed in other DSFGs at z$\sim$1-5. \citep[e.g.,][]{Stacey2010, Gullberg2015}.

\setlength{\tabcolsep}{3pt}
\begin{deluxetable}{lcc}[ht!]
	\tabletypesize{\scriptsize}
	\centering
\tablecaption{Far-IR Fine Structure Lines in PACS/SPIRE Observations \label{tab:fsl}}
\tablehead{
\colhead{Line ID} & \colhead{Rest Wavelength} & \colhead{Line Flux, 3-$\sigma$} \\
\colhead{} & \colhead{($\mu$m)} & \colhead{(10$^{-17}$\,W\,m$^{-2}$)}
}
\startdata
$[$O\,{\sc iv}$]$ ${^2}\rm P_{3/2}\rightarrow {^2}\rm P_{1/2}$ & 25.91 & $<$0.7\tablenotemark{a}  \\
$[$O\,{\sc iii}$]$ ${^3}\rm P_{2}\rightarrow {^3}\rm P_{1}$ & 51.81 & $<$0.6  \\
$[$N\,{\sc iii}$]$ ${^2}\rm P_{3/2}\rightarrow {^2}\rm P_{1/2}$ & 57.34 & $<$0.7  \\
$[$O\,{\sc i}$]$ ${^3}\rm P_{1}\rightarrow {^3}\rm P_{2}$ & 63.18 & $<$0.6  \\
$[$O\,{\sc iii}$]$ ${^3}\rm P_{1}\rightarrow {^3}\rm P_{0}$ & 88.36 & $<$0.8  \\
$[$N\,{\sc ii}$]$ ${^3}\rm P_{2}\rightarrow {^3}\rm P_{1}$ & 121.89 & $<$0.8  \\
$[$O\,{\sc i}$]$ ${^3}\rm P_{0}\rightarrow {^3}\rm P_{1}$ & 145.53 & $<$1.1  \\
$[$C\,{\sc ii}$]$ ${^2}\rm P_{3/2}\rightarrow {^2}\rm P_{1/2}$ & 157.74 & $<$1.4  \\
\enddata
\tablenotetext{a}{Line flux limit from \citet{Wardlow2017}}
\end{deluxetable}

\section{Discussion} \label{sec:Dis}

In order to understand the star formation history and properties of the stellar populations and the interstellar medium of \src, we address the observed properties of the source in the context of synthesized star formation models, and dust emission and spectral energy distribution models.
\pagebreak

\subsection{UV-to-mm Spectral Energy Distribution} \label{subsec:sed}

Leveraging the broad band coverage we have compiled here for \src, we perform SED fitting using CIGALE \citep[Code Investigating GALaxy Emission,][]{Noll2009, Serra2011} and the high-z extension of MAGPHYS \citep[Multi-wavelength Analysis of Galaxy Physical Properties, ][]{Cunha2008, Cunha2015} with all available photometric data points, described in Section \ref{sec:other_obs}. As the search for a lensing galaxy has been unsuccessful \citep{Calanog2014}, it is conceivable that the near-IR emission seen in the HST and IRAC maps is at least in part from the high-z galaxy. As mentioned in Section \ref{subsec:HST}, to estimate the flux density at 1.15 $\mu$m, we only use the emission co-incident with the centroid of the emission seen in SMA observations. As the Spitzer/IRAC beam is big enough to contain emission from the nearby source, for the purpose of SED modeling we assign an error of 50\% on the flux density at 3.6 and 4.5\,$\mu$m.

\begin{figure*}[t!]
\epsscale{0.77}
\plotone{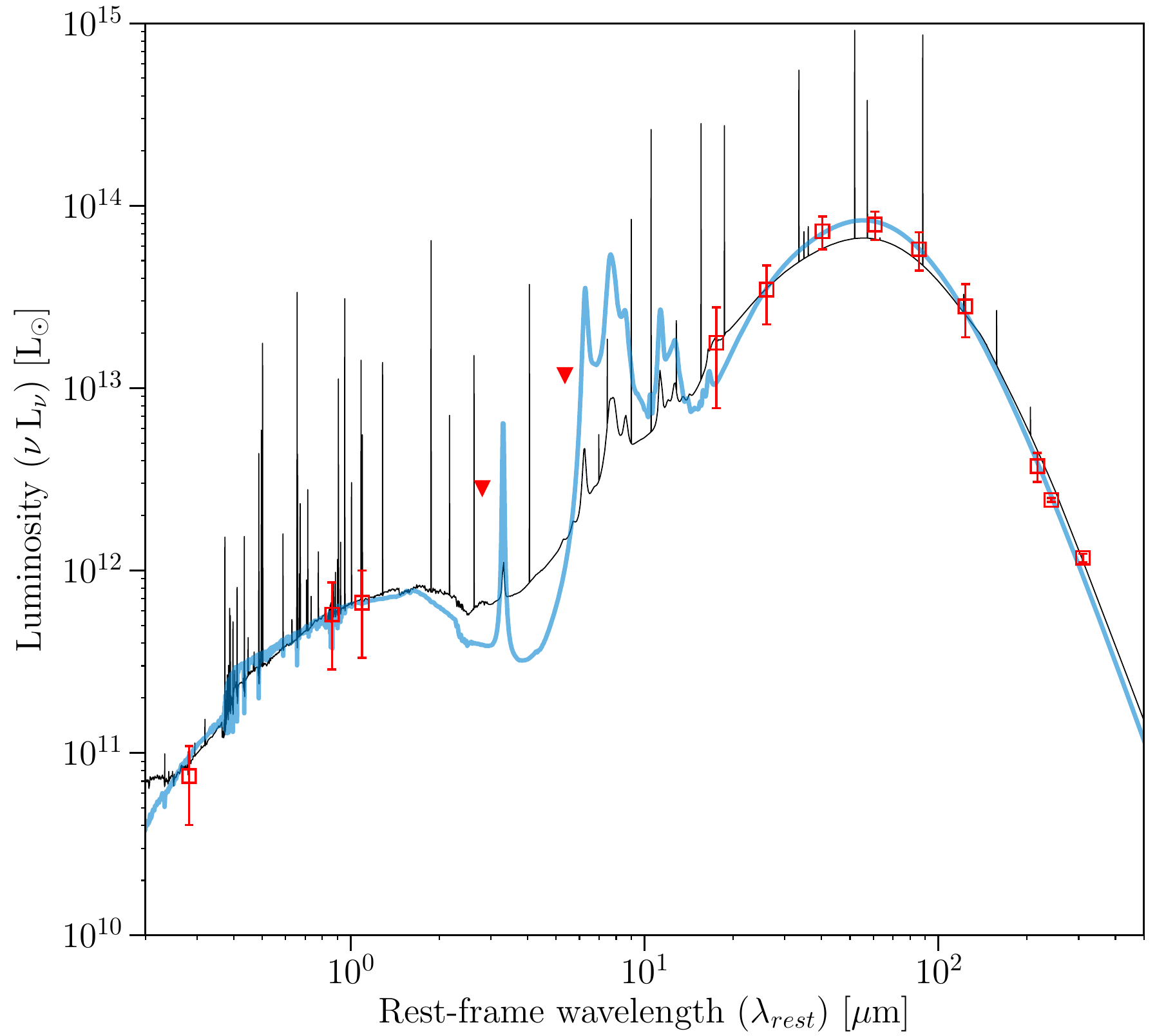}
\caption{Spectral Energy Distribution fit for \src using broadband photometry with CIGALE (black) and MAGPHYS (blue). The red squares are the input photometry measurements with associated errorbars. The red solid triangles indicate the upper limits derived from the WISE W3 and W4 bands.
\label{fig:sed}}
\end{figure*}

CIGALE builds up galaxy SEDs from UV to radio wavelengths assuming a combination of modules. These allow us to model the star formation history (SFH), the stellar emission using population synthesis models \citep[][]{Bruzual2003, Maraston2005}, nebular lines, dust attenuation \citep[e.g.,][]{Calzetti2000}, dust emission \citep[e.g.,][]{Draine2007, Casey2012}, contribution from an AGN \citep[e.g.,][]{Dale2014, Fritz2006}, and radio emission. The SEDs are built while maintaining consistency between UV dust attenuation and far-IR emission from the dust. To model the star formation history, we employ a delayed star formation history prescription used to model high-z star-forming galaxies \citep[e.g.,][]{Ciesla2016} with the dust attenuation from \citet{Calzetti2000}, and the dust emission models from \citet{Draine2007}. Finally, CIGALE performs a probability distribution function analysis for our specified model parameters, and obtains the likelihood-weighted mean value for each parameter.

MAGPHYS uses a Bayesian approach to constrain galaxy-wide physical properties, including the star formation rate, stellar and dust masses, and contributions from both, hot and cold dust components of the ISM. It builds a large library of reference spectra with different star formation histories \citep[using stellar population synthesis models from ][]{Bruzual2003} and dust attenuation properties \citep[using models from ][]{Charlot2000}. Similar to CIGALE, it also ensures energy balance between the optical and UV extinction and the FIR emission. Both CIGALE and MAGPHYS have internal filter libraries, that use instrumental response curves to perform color correction. 

However, the MAGPHYS package does not allow for a possible AGN contribution to the overall SED fit. In order to explore the presence of a hidden AGN in \src, we used the AGN module in CIGALE with templates from \citet{Fritz2006}, to estimate a parameter that constrains the fraction of observed emission that could be due to an AGN. We compare the resulting best-fit models and derived parameters from both CIGALE and MAGPHYS. We find that the AGN contribution is negligible in \src and the observed SED is well explained by a dust-obscured starburst. For such a heavily obscured system, extinction due to a large column of dust may lead to corrections that could be important for deriving physical properties \citep[e.g.,][]{Uzgil2016}. Using a modified blackbody approximation for the dust emission, we estimate the wavelength $\lambda_{0}=c/\nu_{0}$ where the optical depth $\tau_{\nu}=(\nu/\nu_{0})^{\beta}$ reaches unity for \src. In doing so, we make the following assumptions, the resolved dust continuum size from the SMA observations is used to measure the source solid angle, the dust emissivity spectral index $\beta$ is fixed to 1.5, and that the correction due to contrast against the CMB is negligible. We find $\lambda^{rest}_{0}$=32$\pm$8\,$\mu$m and assuming the dust is well mixed in the medium, the correction to the [O\,{\sc iii}] 88\,$\mu$m line luminosity in Section \ref{sec:Z2} would be $\sim 11$\%, well within the reported uncertainty. The SEDs and their best fits from both CIGALE and MAGPHYS are plotted in Figure \ref{fig:sed}. Both SED fits are consistent within the errors and the corresponding best fit parameters are listed in Table \ref{tab:sedprop}. L$_{\rm IR}$ is calculated by integrating under the best-fit SED between $\lambda_{rest}=8-1000\,\mu$m and L$_{\rm FIR}$ by integrating over $\lambda_{rest}=42.5-122\,\mu$m.

\setlength{\tabcolsep}{6pt}
\begin{deluxetable*}{llll}[h!]
	\tabletypesize{\scriptsize}
	\centering
	\tablewidth{\textwidth}
\tablecaption{Derived Physical Properties of \src \label{tab:sedprop}}
\tablehead{
\colhead{Quantity} & \colhead{Value} & \colhead{Unit} & \colhead{Notes/Ref}
}
\startdata
Stellar Mass (M$_{\star}$) & 7.7 $^{+1}_{-4}\,\times10^{10}$ & \mgf\msol & (1) \\
IR Luminosity (L$_{\rm IR}$) & 1.3 $\pm\,0.1\,\times10^{13}$ & \mgf\lsol & (1), (2)\tablenotemark{a} \\
Far-IR Luminosity (L$_{\rm FIR}$) & 8.3 $\pm\,0.9\,\times10^{12}$ & \mgf\lsol & (1), (3)\tablenotemark{b} \\
$[$O\,{\sc iii}$]_{88}$ Luminosity (L$_{\rm OIII}$) & 7.1 $\pm\,0.9\,\times10^{9}$ & \mgf\lsol & (4)\\
Dust Temperature (T$_{\rm dust, MBB}$) & 49.6 $^{+5.6}_{-3.6}$ & K & (1) \\
Dust Temperature (T$_{\rm dust, Draine}$) & 34 $\pm\,1$ & K & (5) \\
Star Formation Rate (SFR) & 916 $^{+88}_{-206}$ & \mgf\msol\,yr$^{-1}$ & (1) \\
Dust Mass (M$_{\rm dust}$) & 5.4 $\pm\,0.9\,\times10^{8}$ & \mgf\msol & (1) \\
Ionized Gas Mass (M$^{\rm H^{+}}_{\rm min}$) & 2.0 $\pm\,0.3\,\times10^{8}$ & \mgf\msol & (4) \\
Ionized Gas Density (n$_{\rm H^{+}}$) & $<$610 & cm$^{-3}$ & (4) \\
Molecular Gas Mass ($\alpha_{\rm CO}$=0.8) & 1.2$\pm\,0.7\,\times10^{10}$ & \mgf\msol & (2) \\
Molecular Gas Mass (Genzel et al. 2015) & 6.1$\pm\,0.7\,\times10^{10}$ & \mgf\msol & (6) \\
\enddata
\tablecomments{(1) This work, parameter derived from best-fit SED with MAGPHYS; (2) \citet{Harris2012}; (3) L$_{\rm FIR}$=8.9$\times10^{13}$\lsol\tablenotemark{b}, \citet{Wardlow2017}; (4) This work, based on [O\,{\sc iii}] 88\,$\mu$m line emission reported in Section \ref{subsec:OIII_prop} and discussion in Section \ref{subsec:SPIRE_spec}; (5) This work, based on U$_{\rm min}$=24, from CIGALE dust emission best-fit model; (6) Based on the method outlined by \citet{Genzel2015} using a variable metallicity dependent $\alpha_{\rm CO}$.}
\tablenotetext{a}{ In both cases the IR luminosity is calculated between 8-1000\,$\mu$m}
\tablenotetext{b}{ L$_{\rm FIR}$ reported by (3) based on luminosity integrated over 40-500\,$\mu$m}
\end{deluxetable*}

\subsection{Constraining the Stellar Population} 

We showed in Section \ref{subsec:OIII_prop} that assuming the [O\,{\sc iii}] line emission we detect in \src arises from H{\sc ii} regions formed by stars, then the young stellar population likely contains $5\times10^{6}\,$\mgf\,upper main sequence stars. In \ref{subsec:sed}, with the help of the broadband SED, we find that indeed the luminosity of the source is dominated by star formation activity and argue against the presence of an AGN. Also, as a part of the SED modeling exercise, we build a stellar population model to constrain the star formation rate and stellar mass. The best fit models suggest a stellar mass content of \src to be $7.7\times10^{10}\,$\mgf\msol with about 12\% of that mass attributed to an ongoing starburst event with an average star formation rate, SFR=916\,\mgf\msol\,yr$^{-1}$. As mentioned in Section \ref{subsec:sed}, the uncertainty of assigning the flux from Spitzer/IRAC photometry to the high redshift source dominates the errors in determining the total stellar mass as reported in Table \ref{tab:sedprop}. We also find that the predicted [O{\sc iii}]\,88$\mu$m line luminosity from the nebular emission component of the best-fit model, L$_{[\rm OIII]}$=(10$\pm$4)$\times$10$^{9}$\,\mgf\lsol is in agreement with the observed line luminosity reported in Section \ref{subsec:OIII_prop}, adding confidence to our interpretation that \src hosts a significant young stellar population.

\pagebreak

Another way to look at the number and mass of the upper main sequence stars is to estimate them using an initial mass function assuming a star formation rate. Here, we assume a Salpeter IMF of the form $\Phi(M)\propto M^{-2.35}$ with an upper mass cutoff of 100\msol and a lower mass cutoff of 1\msol. In a continuous star formation scenario, the equilibrium number for O-stars with mass $>34.4$\,\msol (O5.5 or higher) could be estimated as follows:
\begin{subequations}
      \begin{align}
       \intertext{Total $mass$ of stars formed in an year = SFR,}
       SFR &= \int_{1}^{100}{M\times k\,M^{-2.35}\,dM}\\
       \intertext{Here $k$ is an arbitrary constant of proportionality.}
       \Rightarrow k &= \frac{SFR}{\int_{1}^{100}{M^{-1.35}\,dM}} = \frac{SFR}{2.287}\\
       \intertext{Total $number$ of stars formed per year with ${\rm M>M_{low}}$,}
       {\rm \# \,of\, Stars\, ,N} &= \int_{M_{low}}^{100}{k\,M^{-2.35}}\\
       \intertext{If the main sequence lifetime of star with mass M$_{low}$ is $\tau_{MS}$, then their equilibrium number would be,}
       {\rm N_{Eq}} &\approx \frac{SFR}{2.287}\,\int_{M_{low}}^{100}{M^{-2.35}}\,\times\tau_{MS}
      \end{align}
    \end{subequations}
    
For an O5.5 star, where the main sequence lifetime for the star is $\tau_{MS}$ = 4.9\,Myr \citep{Ekstrom2012}, thus the equilibrium number of O5.5 or higher stars in a continuous star formation scenario with a SFR=916\,\msol\,yr$^{-1}$ would be $\sim (916/2.287)\times4.76\times10^{-3}\times4.9\times10^{6}=9.3\times10^{6}$ or about twice the current number. This simple result demonstrates that the star formation event, if sustained at its present rate, must be less than 5 million years old.

Starburst99  models \citep{Leitherer1999} for continuous star formation utilize the same constraints and provide an independent comparison for the quantities derived here. The Starburst99 models count the number of O-stars as those with T$_{\rm eff} >$30,000 K, which would correspond to all stars above $\sim 15.6\,$\msol \citep{Martins2005}. But, a star of spectral class O5.5 or earlier has a mass M $>34.4\,$\msol, and such stars account for only about 28\% of the total number of stars considered to be O-stars by the Starburst99 models. In the continuous star formation scenario forming 1\,\msol\,yr$^{-1}$, Starburst99 reports 20,800 O-stars after 4.9 Myrs. To get the estimated number of O-stars, $(5\times10^{6}/0.28)\sim1.8\times10^{7},\,$ the Starburst99 models would require an effective star formation rate of 859\,\msol\,yr$^{-1}$, which is consistent with our estimated SFR of 916\,\msol\,yr$^{-1}$ within errors.

\subsection{ Conditions of Interstellar gas and dust} \label{subsec:gd}

Here we will compare the interstellar dust and gas content of \src using available data, particularly gas mass reported by \citet{Harris2012} using $^{12}$CO(1-0) line observations, the dust mass estimate from the SED, and the molecular gas mass estimates using the method described in \citet{Scoville2016} and \citet{Genzel2015}. The continuum measurements on the Rayleigh-Jeans (R-J) tail can be used to independently estimate total molecular gas mass using the method described in \citet{Scoville2016}. The main caveat for using the continuum to derive molecular gas mass estimates using this technique is that the continuum data point should be well on the R-J tail and not near the peak of the dust blackbody emission. Here, we only use the continuum points longward of 890\,$\mu$m (observed frame) or $>210\,\mu$m rest-frame at z=3.127 (see Table \ref{tab:phot}). We use the relationship between total molecular gas mass in the galaxy and continuum flux density observed on the R-J tail by \citet{Scoville2016},

\begin{subequations}
      \begin{align}
       \rm M_{mol} =& 1.78\,\left ( \frac{S_{\nu}}{\rm mJy} \right ) \left (\frac{d_{L}}{\rm Gpc} \right )^{2} \left (\frac{\nu_{obs}}{\nu_{850\,\mu m}} \right )^{-3.8} \left ( \frac{\Gamma_{0}}{\Gamma_{\rm RJ}} \right ) \nonumber\\
       & (1+z)^{-4.8} \,10^{10}\,{\rm M_{\odot}}\, , \label{eq:mmol}\\
      {\rm where,\,}&{\rm \Gamma_{RJ}(z,T_{d},\nu_{obs})} = {\rm \frac{h\nu_{obs}(1+z)/kT_{d}}{e^{h\nu_{obs}(1+z)/kT_{d}}-1}}\, , \label{eq:corr_fac}\\
      {\rm and,\, \Gamma_{0}} &= {\rm \Gamma_{RJ}(0,25\,K,\,\nu_{850\,\mu m}) = 0.7} \nonumber
      \end{align}
    \end{subequations}

It has been often suggested that the temperature of the cold dust component in galaxies derived using modified Blackbody models tends to be biased higher when compared to the expected temperature of the bulk of the dust mass present in the ISM \citep[e.g.,][]{Draine2007, Scoville2016}. As part of the SED modeling we used the dust models from \citet{Draine2007} to constrain the distribution of the ambient interstellar radiation field intensity (U). The models further estimate an average dust temperature based on the minimum intensity of radiation field as T$_{\rm dust}=20\,{\rm U^{1/6}_{min}}$\,K \citep{Draine2010}, which for \src results in T$_{\rm dust}=34\,$K based on U$_{\rm min}=24\pm4$.

In order to be closest to the calibration derived by \citet{Scoville2016}, we use the above dust temperature along with the 896\,$\mu$m continuum observations in Eq(\ref{eq:mmol}), to estimate the molecular gas mass in \src to be M$_{\rm mol}=1.8\times10^{11}\,$\mgf\msol. For comparison, \citet{Harris2012} reported a total molecular gas mass, M$_{\rm mol}=(1.2\pm 0.7)\times10^{10}\,$\mgf\msol based on observations of the $^{12}$CO(1-0) line luminosity and assuming a CO-to-H$_{2}$ conversion factor, $\alpha_{\rm CO}$=0.8\,\msol\,(K-km\,s$^{-1}$\,pc$^{2})^{-1}$. This choice of $\alpha_{\rm CO}$ is typically used for local ULIRGs and to compare molecular mass estimates between high-z SMGs. Even if we use the “luminosity-weighted” dust temperature derived from the SED fit, T$_{\rm dust}=49.6\,$K in Eq(\ref{eq:mmol}), we estimate a molecular gas mass of $1.3\times10^{11}\,$\mgf\msol. We also verify that we get similarly high molecular gas mass estimates using the continuum data points reported by \citet{Yang2016} at 1 and 1.25\,mm (240 and 300$\,\mu$m respectively in the rest frame of \src). We apply the appropriate correction factors ${\rm \Gamma_{RJ}}$ as defined in Eq(\ref{eq:corr_fac}), that accounts for deviation from the default calibration at 850$\,\mu$m due to redshift and dust temperature, and estimate the molecular gas mass as $(1.6-2)\times10^{11}\,$\mgf\msol. Therefore, using the submm dust continuum method outlined by \citet{Scoville2016}, we find the estimated molecular gas mass to be 10-16 times larger than the molecular gas mass derived using the $^{12}$CO(1-0) line observations with a ULIRG-like conversion factor of 0.8\,\msol\,(K-km\,s$^{-1}$\,pc$^{2})^{-1}$.

We derive a total dust mass of, M$_{\rm dust}=(5.4\pm1)\times10^{8}\,$\mgf\msol from the best-fit SED model. \citet{Genzel2015} provide a relationship to estimate the dust mass using star formation rate and modified dust blackbody temperature, 

\begin{equation} \label{eq:genzel_md}
{\rm M_{dust}=1.2\times 10^{15}\,}\,\left (\frac{\rm SFR}{\rm M_{\odot}\,yr^{-1}} \right )\,\left ( \frac{\rm T^{MBB}_{dust}}{\rm K} \right )^{-5.5},
\end{equation}

Using, T$_{\rm dust}$=49.6\,K and SFR=916\,\msol\,yr$^{-1}$ in Eq(\ref{eq:genzel_md}), we find M$_{\rm dust}$=5.1$\times10^{8}\,$\msol which is consistent with the total dust mass derived from the best fit SED. Now, we compare the dust-to-gas mass ratio using the different molecular gas mass estimates with a metallicity dependent dust to gas ratio from \citet{Leroy2011},

\begin{equation} \label{eq:leroy_gdr}
\log_{10} \delta_{dg} = \log_{10} \left ( \frac{\rm M_{dust}}{\rm M_{mol}}\right ) = -2 + 0.85\times(12+\log_{10}[{\rm O/H}]-8.67)\, ,
\end{equation}

If we use the $^{12}$CO(1-0) measurement along with a ULIRG like conversion factor, we find that \src has a metallicity, 12+log[O/H]=9.4, whereas, using the estimate of the molecular gas mass using the R-J continuum, we find a metallicity, 12+log[O/H]=8.2. In contrast, using a fitting function for metallicity by combining the stellar mass-metallicity relation at different redshifts as given in Eq(12) by \citet{Genzel2015}, we find a metallicity, 12+log[O/H]=8.6$\pm$0.1. Using this estimate of metallicity in Eq(\ref{eq:leroy_gdr}), we find a $\delta_{dg}$=(0.8-1)$\times 10^{-2}$. This would imply a total molecular gas mass of M$_{\rm mol}$=6.1$\pm\,0.7\times 10^{10}\,$\mgf\msol and an implied $\alpha_{\rm CO}\approx4\,$\msol\,(K-km\,s$^{-1}$\,pc$^{2})^{-1}$, quite similar to that observed in the Milky Way. 
For a galaxy undergoing a vigorous starburst event of the likes we suggest for \src, the filling fraction of star-forming, denser, gas could be higher than the typical local ULIRG values. \citet{Downes1993} parametrized the conversion factor, $\alpha_{\rm CO} \propto \sqrt{n}/T_{b}\,$, where n is the the average H$_{\rm 2}$ density of the gas clouds and $T_{b}$ is the intrinsic brightness temperature of the $^{12}$CO(1-0) line. A higher volume-averaged density of the medium could account for the relatively high $\alpha_{\rm CO}$ value suggested here. We use this estimate of the molecular gas mass for further discussion.

Even though the various molecular gas mass estimates vary significantly, the implied baryonic gas fraction $\left ( {\rm f_{gas}} = \frac{\rm M_{mol}}{\rm M_{mol}+M_{stellar}} \right )$, is f$_{\rm gas}$=0.44$^{+0.23}_{-0.30}$, is similar to observed gas fractions in gas-rich high-z SMGs \citep[e.g.,][]{Tacconi2013, Tacconi2017}. The high gas fraction and yet a relatively short gas depletion time-scale of 66 Myrs, along with the enormous star formation rate, agree with our understanding of the star formation activity in \src. Specifically, that it is undergoing a star formation episode building up its stellar mass and is currently above the star-forming main sequence at z$\sim$3 \citep{Speagle2014}.

\subsection{Is \src strongly lensed?} \label{subsec:lensing}

With the high signal–to–noise, SMA continuum observations (FWHM: 0\farcs93$\times$0\farcs63), we derive the de-convolved source size of the dust emitting disk as (0\farcs7$\pm$0\farcs1)$\times$(0\farcs6$\pm$0\farcs1), which corresponds to a physical size of (5.6$\times$4.5)\,kpc or r$_{\rm dust} \sim $ 5\,kpc. In terms of area, this is about 7 times larger than the typical size of the dust disk, r$_{\rm dust} \sim $1.8\,kpc, found by \citet{Hodge2016} in a resolved study of sixteen z$\sim$2.5 SMGs. As gravitational lensing spreads the intrinsic source over a larger area, a lensing magnification factor of $\mu\sim7$ could make \src's dust disk consistent with those observed in other high-z SMGs.  

The observed line widths for spectral lines tracing various transitions of CO and H$_{2}$O in \src are about $\sim$225\,km\,s$^{-1}$. As a comparison, in a study of J$>$2 CO line observation in DSFGs presented in \citet{Bothwell2013}, the mean value for CO line widths was found to be (510$\pm$80)\,km\,s$^{-1}$. This might indicate that \src is either an almost face-on disk or not an intrinsically massive galaxy. As the CO line emission has been used to study the mass and kinematics of galaxies, \citet{Harris2012} suggested that there should be a ``Tully-Fisher''-like relationship between the CO line width and the intrinsic line luminosity. Based on the bright $^{12}$CO(1-0) line detection and the small line width they suggested a lensing magnification factor $\mu\sim17\pm11$ for \src.

We can also estimate the CO emitting size, even without spatially resolved observation, assuming that the CO emission is optically thick. Following Eq(2) from \citet{Solomon2005},

\begin{equation}
L^{'}_{\rm CO} = 1.13\,(T_{\rm ex} - T_{\rm CMB}(z))\,\Delta v_{\rm FWHM}\,r_{\rm CO}^{2}
\end{equation}

The temperature of the CMB at z=3.127 is T$_{\rm CMB}$ =11.2\,K. The first rotational transition of the $^{12}$CO line only lies about 5.51K above ground and the molecules can be easily excited by the ambient radiation field and by collisions in the gas. We apply an excitation temperature of T$_{\rm ex}$=30\,K, assuming that it should be higher than the background CMB temperature and perhaps lower than the the dust temperature, T$_{\rm dust}$=50\,K. Now, with the observed line luminosity, L$^{'}_{\rm CO}$=1.5$\times10^{11}$ K-km\,s$^{-1}$\,pc$^{2}$ and measured line FWHM, $\Delta v_{\rm FWHM}$=225\,km\,s$^{-1}$, we derive a r$_{\rm CO}$ = 5.6\,kpc, which is about the same as the size of the dust emitting region derived above. We note that the CO emitting size estimate is weakly dependent on our choice of T$_{\rm ex}$.

For a simple rotating disk model following \cite{Neri2003}, the dynamical mass of the system can be estimated as,

\begin{equation}
{\rm M_{dyn}\,sin^{2}i}=2.33\times10^{5}\,\left (\frac{\rm \Delta V}{\rm km\,s^{-1}} \right )^{2}\,\left (\frac{\rm r}{\rm kpc} \right )\,{\rm M_{\odot}}\,,
\end{equation}

where i=cos$^{-1}$($\frac{4.5}{5.6}$)=36.5$\degr$ is the inclination angle estimated from the resolved dust continuum, {\rm $\Delta$V} is the CO line velocity dispersion in km s$^{-1}$ (=FWHM/2$\sqrt{2\,\ln 2}$), and r is the disk radius in kpc as derived above. We estimate the dynamical mass of the galaxy to be, M$_{\rm dyn}$=3.4$\times10^{10}\,$\msol. Since this estimate of the dynamical mass is atleast 2 times smaller than either our estimates for the stellar mass or the gas mass, our assumption that the observed ellipticity reflects inclination is suspect. In order to make a fair comparison between the estimates of dynamical mass, molecular gas mass, and stellar mass, we need to consider how the lensing magnification factor ($\mu$) and the choice of CO-to-H$_{2}$ conversion factor ($\alpha_{\rm CO}$) affect our measurements. We expect the dynamical mass of the system to be larger than the estimates of either the stellar mass or the molecular gas mass. The apparent spatial size scales as $r \propto\sqrt{\mu}$, but the apparent luminosity (CO and stars) scales as $L\propto\mu$. The choice of the CO-to-H$_{2}$ conversion factor ($\alpha_{\rm CO}$) only changes the estimate for the molecular gas mass, $M_{\rm mol}\propto\alpha_{\rm CO}$. If we assume the source is unlensed (largest r, M$_{\rm dyn}$) and use $\alpha_{\rm CO}$=0.8 (lowest estimate of the molecular gas), this still results in M$_{\rm mol}\sim$ 3.5$\times$M$_{\rm dyn}$. Without changing the $\alpha_{\rm CO}$ (increasing $\alpha_{\rm CO}$, widens the discrepancy), in order to match the dynamical mass to the molecular gas mass, we would need to increase the magnification factor to $\mu$=12.5. A less inclined orientation would bring the estimates closer together, but a simpler explanation for the observed line and continuum luminosities could be amplification due to the affect of gravitational lensing.

\section{Results \& Conclusions} \label{sec:results}
We have presented sensitive ground based THz spectroscopy with ZEUS-2 at the APEX Telescope, detecting the [O{\sc iii}]\,88\,$\mu$m line in \src, a high redshift sub-millimeter galaxy at z=3.127. The luminosity in the line is, 7.1$\times 10^{9}\,$\mgf\lsol, which indicates the presence of a large number of upper main sequence O-stars and allows us to constrain the number of ionizing photons available. In the high density, high temperature limit, we derive that the minimum mass of ionized gas required to support the observations would be 2.8$\times10^{8}\,$\mgf\msol or about 0.33\% of the estimated total molecular gas, M$_{\rm mol}=6.1\times10^{10}\,$\mgf\msol.

The upper limit from the SPIRE spectrum on the [C{\sc ii}]/FIR ratio, $\leq$0.4\% would be consistent with a compact starburst, but the small velocity width and single gaussian profile in CO and H$_{2}$O lines argue against a major merger as the source of the observed [C{\sc ii}]/FIR line to continuum ratio. We also constrain the ionized gas density to be, n$_{\rm H^{+}}<$610\,cm$^{-3}$ using the limit for the [O{\sc iii}]\,52\,$\mu$m line from the SPIRE spectrum along with our detection of the [O{\sc iii}]\,88\,$\mu$m line. 

Using rest-frame UV to mm wavelength observations, we constrain a broadband SED for the source and rule out any significant contribution by an obscured AGN. We find, using both the SED and extrapolating the equivalent number of 5$\times10^{6}\,$\mgf\,O5.5-O3 stars estimated using the  the observed [O{\sc iii}] line emission, that the on-going star formation event in \src has contributed up to 0.55-1$\times10^{10}\,$\mgf\msol to its stellar mass, i.e., up to 7-12\% of the total stellar mass, M$_{\rm stellar}$=7.7$\times 10^{10}\,$\mgf\msol. 

A gas fraction of f$_{\rm gas}$=0.44$^{+0.23}_{-0.30}$, indicates that the galaxy has an abundant supply of gas to sustain star-formation over the next 66 million years, effectively doubling its stellar mass and end up on the star-forming main sequence at z$\sim$3. 

We also detect rest-frame near-UV emission in the 1.1\,$\mu$m HST/WFC3 image, consistent with the centroid of the dust emission seen in the SMA 896\,$\mu$m continuum image, it appears patchy and there is another source $\sim$1\arcsec\,away along the NE-SW axis but no obvious signature of gravitational lensing like extended arcs or an Einstein ring are seen. The apparent brightness of the galaxy suggests that it could be lensed and the low dynamical mass estimate is consistent with the lensing scenario. Higher resolution observations are required to reveal the true nature of this otherwise apparent gargantuan, specifically to understand the role gravitational lensing plays in amplifying the observed emission and uncovering the intrinsic nature of \src.

\acknowledgments
The authors would like to thank the anonymous referee for a constructive report and suggestions that helped improve the quality of the manuscript. ZEUS-2 development and observations are supported by NSF grants AST-0705256, AST-0722220, AST-1105874, and AST-1109476, and a grant from Georgia Southern University. This publication is based on data acquired with the Atacama Pathfinder Experiment (APEX) Telescope. APEX is a collaboration between the Max-Planck-Institut f{\"u}r Radioastronomie, the European Southern Observatory, and the Onsala Space Observatory. The authors would like to thank the APEX staff whose excellent support helped to make this work possible. The Herschel-ATLAS is a project with Herschel, which is an ESA space observatory with science instruments provided by European-led Principal Investigator consortia and with important participation from NASA. Part of this work is based on observations made with the NASA/ESA Hubble Space Telescope, obtained from the data archive at the Space Telescope Science Institute. STScI is operated by the Association of Universities for Research in Astronomy, Inc. under NASA contract NAS 5-26555. Part of this work is also based on observations made with the Spitzer Space Telescope, which is operated by the Jet Propulsion Laboratory, California Institute of Technology under a contract with NASA.

\vspace{5mm}
\textit{Facilities:} APEX(ZEUS-2), HST(WFC3), Spitzer(IRAC), WISE, Herschel(PACS and SPIRE), SMA





\end{document}